\pdfoutput=1
\RequirePackage{ifpdf}
\ifpdf % We are running pdfTeX in pdf mode
\documentclass[pdftex]{sigma}
\else
\documentclass{sigma}
\fi

\begin{document}

\allowdisplaybreaks

\renewcommand{\thefootnote}{$\star$}

\renewcommand{\PaperNumber}{052}

\FirstPageHeading

\ShortArticleName{Invariant Discretization Schemes Using Evolution--Projection Techniques}

\ArticleName{Invariant Discretization Schemes \\
Using Evolution--Projection Techniques\footnote{This paper is a~contribution to the Special Issue  ``Symmetries of Dif\/ferential
Equations:  Frames, Invariants and~Applications''.
The full collection is available at
\href{http://www.emis.de/journals/SIGMA/SDE2012.html}{http://www.emis.de/journals/SIGMA/SDE2012.html}}}

\Author{Alexander BIHLO~$^{\dag\ddag}$ and Jean-Christophe NAVE~$^\ddag$}

\AuthorNameForHeading{A.~Bihlo and J.-C.~Nave}

\Address{$^\dag$~Centre de recherches math\'{e}matiques, Universit\'{e} de Montr\'{e}al,\\
 \hphantom{$^\dag$}~C.P.\ 6128, succ.\ Centre-ville, Montr\'{e}al (QC) H3C 3J7, Canada}
\EmailD{\href{mailto:bihlo@crm.umontreal.ca}{bihlo@crm.umontreal.ca}}

\Address{$^\ddag$~Department of Mathematics and Statistics, McGill University,\\
 \hphantom{$^\ddag$}~805 Sherbrooke W., Montr\'{e}al (QC) H3A 2K6, Canada}
\EmailD{\href{mailto:jcnave@math.mcgill.ca}{jcnave@math.mcgill.ca}}

\ArticleDates{Received September 27, 2012, in f\/inal form July 28, 2013; Published online August 01, 2013}

\Abstract{Finite dif\/ference discretization schemes preserving a subgroup of the maximal Lie invariance group of the one-dimensional linear heat equation are determined. These invariant schemes are constructed using the invariantization procedure for non-invariant schemes of the heat equation in computational coordinates. We propose a new methodology for handling moving discretization grids which are generally indispensable for invariant numerical schemes. The idea is to use the invariant grid equation, which determines the locations of the grid point at the next time level only for a single integration step and then to project the obtained solution to the regular grid using invariant interpolation schemes. This guarantees that the scheme is invariant and allows one to work on the simpler stationary grids. The discretization errors of the invariant schemes are established and their convergence rates are estimated. Numerical tests are carried out to shed some light on the numerical properties of invariant discretization schemes using the proposed evolution--projection strategy.}

\Keywords{invariant numerical schemes; moving frame; evolution--projection method; heat equation}

\Classification{65M06; 58J70; 35K05}

\section{Introduction}

Discretization schemes for dif\/ferential equations that are not solely constructed for the sake of reducing the local discretization error as much as possible, but rather to preserve some of the intrinsic properties of these dif\/ferential equations have become increasingly popular over the last decades. While preserving one of these properties, namely conservation laws, led to the design of conservative discretization schemes which are quite popular in the scientif\/ic community~\cite{brid06Ay,hair06Ay,leim04Ay} (and in particular in the geosciences, e.g.~\cite{fran02Ay,somm09Ay}), there are other geometric features of dif\/ferential equations that can be attempted to be preserved as well that have received less attention from the side of numerical analysis so far. One of these features are symmetries of dif\/ferential equations. While there have been theoretical advancements on the methodologies of f\/inding numerical schemes that preserve the maximal Lie invariance groups of systems of dif\/ferential equations over the past 20 years or so~\cite{doro11Ay,kim08Ay,rebe11Ay,vali05Ay}, little is known about the numerical properties of these invariant schemes. A part of the problem is that while conservation laws are always properties of the solutions of a dif\/ferential equation, symmetries are by def\/inition properties of dif\/ferential equations. Therefore, it is a standing question whether a discretization scheme that preserves numerically a property of a dif\/ferential equation also improves the quality of the numerical solution of that discretized dif\/ferential equation.

The present paper is devoted to an investigation of this question and related problems exempli\-f\/ied with invariant discretization schemes for the linear heat equation. The heat equation is particularly suited for this investigation as it is a canonical example in the group analysis of dif\/ferential and dif\/ference equations. Moreover, there are already several studies devoted to invariant numerical schemes for this equation~\cite{baki97Ay,doro03Ay,vali05Ay}. At the same time, in none of these \mbox{existing} works a~deeper background analysis of the numerical properties (e.g.\ order of ap\-proxima\-tion or stability) of the developed schemes was investigated. A f\/irst account on numerical properties of invariant numerical schemes for the heat equation was given in~\cite{dawe08Ay}, in which a~numerical comparison of invariant and non-invariant schemes for the heat equation regarding accuracy was presented.

There are several reasons why less attention has been paid so far on the numerical analysis of invariant schemes (with the exception of the works~\cite{chha10Ay,dawe08Ay}). One of the reasons is that the f\/ield of invariant discretization schemes is still in its early stages, with new conceptual algorithms being developed only recently~\cite{bihl12By,budd01Ay,budd96Ay,doro11Ay,kim06Ay,kim08Ay,levi06Ay,rebe11Ay}. Another reason is that invariant f\/inite dif\/ference schemes generally require the use of \textit{adaptive moving meshes}, i.e.\ it is necessary to include a non-trivial mesh equation in the discretization problem. Moving meshes lead to non-uniform grid point distributions and, in the multi-dimensional case, to non-orthogonal grids. The analysis of schemes on such meshes is considerably more dif\/f\/icult than that for related dif\/ference schemes on f\/ixed, uniform and orthogonal meshes. Due to this second reason, most invariant numerical schemes so far have been constructed only for (1+1)-dimensional single evolution equations, as in that case moving meshes can be handled still with limited ef\/fort. Although we will be concerned with a (1+1)-dimensional equation in the present paper too, the methods used subsequently can be employed in the multi-dimensional case without substantial modif\/ication.

The new approach we propose here is to use the invariant grid equation only for a single time step and then to interpolate the solution to the regular grid. The important observation is that this interpolation can be done in an invariant way, i.e.\ projecting the solution does not break the invariance of the scheme itself. At the same time, the possibility to project the solution of an invariant scheme to a regular grid is highly desirable as in the multi-dimensional case a freely evolving grid can cause severe numerical problems. Moreover, for realistic numerical models, as e.g.\ employed in weather and climate predictions, it is in general hard to use adaptive meshes as the discretization of the governing equations is only one part of such model. Other parts are related to the numerical data assimilation, i.e.\ the preparation of the initial conditions for the numerical model and this step usually involves the forecasting model itself. As the assimilation of the initial conditions cannot be done on an evolving mesh (because the data are given at f\/ixed locations only) this at once renders invariant schemes on moving meshes impractical. Equally important, any realistic numerical model for a nonlinear system of partial dif\/ferential equations has to contain subgrid-scale models, which mimic the ef\/fects of processes taking place at those scales that the numerical model is not capable of resolving explicitly~\cite{sten07Ay,stul88Ay}. The construction of subgrid-scale models for non-resolved physical processes involves in general ad-hoc arguments and these arguments rely on the particular scale on which the unresolved processes take place. As a moving mesh locally changes the resolution and thus impacts what processes are explicitly resolved by a numerical model, subgrid-scale schemes have to be designed that can operate on grids with varying resolution. For realistic processes (which are usually not self-similar), this might be dif\/f\/icult to achieve in practice.

All what was said above objects against invariant numerical schemes for multi-dimensional systems of dif\/ferential equations using freely evolving meshes. Thus, whether mathematically feasible or not, such schemes would be of less practical interest. This is why other approaches should be sought that on the one hand allow one to retain the invariance group of a system of dif\/ferential equations in the course of discretization and on the other hand yield schemes that are practical to avoid the above mentioned and related problems. The proposed invariant \emph{evolution--projection strategy} we are going to introduce below may be considered as one such approach.

The further organization of this article is as follows. The subsequent Section~\ref{sec:ConstructionOfInvariantSchemes} features a summary and some extensions on the various methods to construct invariant discretization schemes. In Section~\ref{sec:HeatEquation} the heat equation along with its maximal Lie invariance group $G$ is presented. It is discussed which subgroup $G^1$ of $G$ we aim to present when constructing invariant numerical discretization schemes. The selection of $G^1$ is based on preserving the class of periodic boundary value problems we are focussing on. Section~\ref{sec:DifferentialInvariantsHeatEquation} contains the construction of an equivariant moving frame for $G^1$ along with a presentation of some lower order dif\/ferential invariants of $G^1$. In Section~\ref{sec:InvariantDiscretizationHeatEquation} invariant discretization schemes for the heat equation in computational coordinates are found. The local discretization errors of these schemes are established in Section~\ref{sec:NumericalPropertiesSchemesHeatEquation}. In Section~\ref{sec:InvariantInterpolationSchemes} we introduce the new idea of invariant interpolation schemes that can be used to project the numerical solution obtained from an invariant scheme on a moving mesh to the regular grid. The numerical analysis as well as some numerical tests for the schemes proposed in this paper are found in Section~\ref{sec:NumericalTests}. The summary of this article is presented in the f\/inal Section~\ref{sec:ConculsionsHeatEquation}.

\section{Construction of invariant discretization schemes}\label{sec:ConstructionOfInvariantSchemes}

The construction of invariant discretization schemes for dif\/ferential equations can be seen as a~part of the ongoing ef\/fort to turn group analysis into an ef\/f\/icient tool for the analysis of dif\/ference equations, see e.g.\ the review article~\cite{levi06Ay}. As of now, there are three main methods that are used to construct invariant discretization schemes.

\subsection{Dif\/ference invariant method}\label{sec:ConstructionOfInvariantSchemesDordonitsyn}

The f\/irst method was developed by V.\ Dorodnitsyn, see~\cite{baki97Ay,doro11Ay,doro03Ay,levi06Ay,vali05Ay}. It uses the inf\/ini\-te\-si\-mal generators of one-parameter symmetry groups of the system of dif\/ferential equations under consideration that span the maximal Lie invariance algebra $\mathfrak g$ of this system. These generators are of the form
\[
 \mathrm{v}=\zeta^j(x,u)\partial_{x^j}+\eta^\alpha(x,u)\partial_{u^\alpha}=\zeta(x,u)\partial_x+\eta(x,u)\partial_u,
\]
where $x=(x^1,\dots,x^p)$ and $u=(u^1,\dots,u^q)$ are the tuples of independent and dependent variables, respectively. Here and in the following, the summation convention over repeated indices is used. Rather than prolonging $\mathrm{v}$ to higher order derivatives of $u$ with respect to $x$, which is standard in the symmetry analysis of dif\/ferential equations~\cite{blum89Ay,olve86Ay,ovsi82Ay}, in this method the vector f\/ields are prolonged to all the points of the \emph{discretization stencil}, i.e.\ the collection of grid points which are necessary to approximate a given system of dif\/ferential equation up to a~desired order. This prolongation is of the form
\[
 \mathrm{pr}\, \mathrm{v}=\sum_{i=1}^m\zeta(x_i,u_i)\partial_{x_i}+\eta(x_i,u_i)\partial_{u_i},
\]
where $x_i=(x^1_i,\dots,x^p_i)$ and $u_i=(u^1_i,\dots,u^q_i)$, i.e.\ it is done by evaluating the vector f\/ield $\mathrm{v}$ at all $m$ stencil points $z_i=(x_i,u_i)$ and summing up the result. An example for such a prolongation is given in Remark~\ref{rem:CheckingInvarianceOfScheme} in Section~\ref{sec:InvariantDiscretizationHeatEquation}.

As a next step, the invariants of the group action are found by invoking the inf\/initesimal invariance criterion~\cite{doro11Ay,olve86Ay}, which in the present case is $\mathrm{pr}\,\mathrm{v}(I)=0$. The functions $I$ that fulf\/ill this condition for all $\mathrm{v}\in\mathfrak g$ are termed \textit{difference invariants}.

Once the dif\/ference invariants on the stencil space are found, one then has to assemble these invariants together to a f\/inite dif\/ference approximation of the given system of dif\/ferential equations. By construction, this procedure guarantees that the resulting numerical scheme is invariant under the symmetry group of the original system of dif\/ferential equations.

The main drawback of this method is that it might be hard to f\/ind a combination of dif\/ference invariants that approximates a system of dif\/ferential equations in the multi-dimensional case. The problem is, as discussed in the introduction, that invariant schemes generally require the use of moving and/or non-orthogonal grids. Formulating consistent discretization schemes using dif\/ference invariants as building blocks on moving meshes is rather challenging in higher dimensions and thus limited the application of this method to the case of single $(1+1)$-dimensional evolution equations. We stress though that this problem only enters at the stage of combining dif\/ference invariants to a discretization scheme. Computing dif\/ference invariants in the multi-dimensional case can be done as ef\/fectively as computing dif\/ferential invariants for multi-dimensional problems using inf\/initesimal techniques.

\subsection{Invariant moving mesh method}\label{sec:ConstructionOfInvariantSchemesMovingMesh}

Retaining the invariance of f\/inite dif\/ference schemes under the maximal Lie invariance groups of physically relevant time-dependent dif\/ferential equations often requires the use of moving meshes. This is true both for the f\/inite dif\/ference method discussed in the previous Section~\ref{sec:ConstructionOfInvariantSchemesDordonitsyn} and the moving frame method to be discussed in the next Section~\ref{sec:ConstructionOfInvariantSchemesMovingFrame}. This kind of mesh adaptation in which the number of grid points remains constant throughout the integration is referred to as $r$-adaptivity in the f\/ield of adaptive numerical schemes~\cite{budd09Ay,huan10By}.

The standard strategy to handle $r$-adaptive meshes is to regard the grid adaptation as a time-dependent mapping from a f\/ixed reference space of \emph{computational coordinates} to the physical space of the independent variables of the dif\/ferential equation, i.e.\ $x=x(\xi)$ for $\xi=(\xi^1,\dots,\xi^p)$ being the computational variables. Without loss of generality, we assume that $\xi^1=\tau=t$ is the time variable. The dependent variables $u$ are expressed in the computational space by setting $\bar u(\xi)=u(x(\xi))$. For the sake of simplicity we will omit the bars henceforth.

The signif\/icance of the computational coordinates is to provide a reference frame that remains stationary and orthogonal even in the presence of grid adaptation in the physical space of coordinates. In the course of discretization the variable $\xi$ labels the position of the grid points in the mesh and this labeling stays unchanged during the mesh adaptation. Thus, the computational variables can be interpreted physically as Lagrangian coordinates and their invariance under the motion of the grid is equivalent to the identity of f\/luid particles in ideal hydrodynamics.

Because by construction the grid remains orthogonal in the $\xi$-coordinates, the usual f\/inite dif\/ference approximations for derivatives can be used in the space of computational variables. This simplif\/ies both the practical implementation of the discretization method and the numerical analysis of the resulting schemes.

The expression of the initial physical system of dif\/ferential equations in terms of computational variables leads to a system of equations that explicitly includes the mesh velocity $x_\tau$, which is yet to be determined in order to close the resulting numerical scheme. A prominent strategy for determining the location of the grid points at the subsequent time level in the one-dimensional case uses the \emph{equidistribution principle}, which in its dif\/ferential form is $(\rho x_\xi)_\xi=0$, where $\rho$ is a monitor function that determines the areas of grid convergence and divergence. For higher-dimensional problems, equidistribution has to be combined with heuristic arguments, see~\cite{huan10By} for more details.

The invariance of the initial dif\/ferential equations is brought into the scheme by adequately specifying the monitor function $\rho$. In~\cite{budd96Ay} it was pointed out that using monitor functions that preserving the scale-invariance of a dif\/ferential equation is particularly relevant in cases where the equation is capable of developing a blow-up solution in f\/inite time, see also~\cite{budd09Ay,budd99Ay,huan10By}. This f\/inding is generalized upon requiring that the monitor function is chosen in a manner such that the equidistribution principle is invariant under the same symmetry group as the original dif\/ferential equation. For a number of symmetry groups this appears to be possible, see~\cite{bihl12By} for an example.

The invariant moving mesh method was recently extended in~\cite{bihl12By}. The idea of this extension is to transform the initial system of dif\/ferential equations to the space of computational coordinates and to determine the form of the symmetry transformations in the computational space. The equations in the computational space are then discretized such that the resulting scheme mimics the transformation behavior of the continuous case. The main advantage of this approach is that it allows one to retain an initial conserved form of the system of dif\/ferential equations and thus to numerically preserve certain conservation laws in the invariant scheme. This is relevant as preserving conservation laws in the course of invariant numerical modeling is yet a pristine problem. An exception to this is the discretization of equations that follow from variational principles, which, if done in a proper way, can lead to the simultaneous preservation of both symmetries and associated conservation laws, owing to the discrete Noether theorem. See, e.g.~\cite{budd01Ay} for an example of such an invariant Lagrangian discretization.

Another advantage of the extension proposed in~\cite{bihl12By} is that it allows one to f\/ind invariant numerical schemes without the detour of dif\/ference invariants. This is essential as it can happen that the single equations in a system of dif\/ferential equations cannot be approximated directly in terms on dif\/ferential invariants but only in combination with other equations of that system. If this is the case it is not natural to attempt to discretize the system using dif\/ference invariants as this would lead to rather cumbersome discretization schemes.

\subsection{Moving frame method}\label{sec:ConstructionOfInvariantSchemesMovingFrame}

The third method is the most recent one~\cite{chha10Ay,chha11Ay,kim06Ay,kim08Ay,kim04Ay,rebe11Ay}. It relies on the notion of \textit{equivariant moving frames} and their important property to provide a mapping that allows one to associate an invariant function to any given function. As we will mostly work with this method in the present paper, we describe it in greater detail here. We collect some important notions on moving frames below, a more comprehensive exposition can be found in the original references~\cite{cheh08Ay,fels98Ay,fels99Ay,olve01Ay,olve07Ay,rebe11Ay}.

\begin{definition}
 Let $G$ be a f\/inite-dimensional Lie group acting on a manifold $M$. A \emph{$($right$)$ moving frame} $\rho$ is a smooth map $\rho\colon M\to G$ satisfying the equivariance property
 \begin{gather}\label{eq:DefinitionMovingFrame}
  \rho(g\cdot z)=\rho(z)g^{-1},
 \end{gather}
 for all $z\in M$ and $g\in G$.
\end{definition}

\begin{theorem}
 A moving frame exists in the neighborhood of a point $z\in M$ if and only if the group~$G$
 acts freely and regularly near~$z$.
\end{theorem}

\emph{Local freeness} of a group action means that $\tilde z=g\cdot z=z$ for all $z$ from a suf\/f\/iciently small neighborhood of each point on $M$ only holds for $g$ being the identity transformation, which implies that all the group orbits have the same dimension. Here and throughout the paper, a~tilde over a variable denotes the corresponding transformed form of that variable. \emph{Regularity} of a group action requires that there exists a~neighborhood for each point $z\in M$, which is intersected by the orbits of~$G$ into a pathwise connected subset.

When a group $G$ does not act freely on $M$, its action can be made free upon extending it to a~suitably high-order \emph{jet space} $J^n=J^n(M,p)$ of $M$, $0\le n\le\infty$. Locally, the $n$th order jet space of a $p$-dimensional submanifold of $M$ has coordinates $z^{(n)}=(x,u^{(n)})$, where as in the previous subsections $x=(x^1,\dots,x^p)$ are considered as the independent variables, $u=(u^1,\dots,u^q)$, $q=\dim M-p$, are the dependent variables and $u^{(n)}$ collects all the derivatives of $u$ with respect to $x$ of order not greater than $n$ including $u$ as the zeroth order derivatives. In practice, the prolongation of the group action of $G$ on $J^n$ is implemented using the chain rule.

Moving frames are determined using a \emph{normalization} procedure. The steps to f\/ind a moving frame for a group action $G$ are the following: (i)~Def\/ine a \emph{cross-section} to the group orbits. A~cross-section $C$ is any submanifold $C\subset M$ of complementary dimension to the dimension $r$ of the group orbits, i.e.\ $\dim C=\dim M-r$, that intersects each group orbit once and transversally. Usually coordinate cross-sections are chosen in which some of the coordinates of~$M$ (or of~$J^n$ if the group action is not free on $M$) are set to constants, i.e.\ $z^1=c^1,\dots,z^r=c^r$. (ii)~The algebraic system $\tilde z^1=(g\cdot z)^1=c^1,\dots,\tilde z^r=(g\cdot z)^r=c^r$ is solved for the group parameters $g=(g_1,\dots,g_r)$. The resulting expression $g=\rho(z)$ is the moving frame.

Moving frames can be used to map any given function to an invariant function by a procedure called \emph{invariantization}.

\begin{definition}
 The \emph{invariantization} of a real-valued function $f\colon M\to \mathbb{R}$ using the (right) moving frame $\rho$ is the function $\iota(f)$, which is def\/ined as $\iota(f)(z)=f(g\cdot z)|_{g=\rho(z)}=f(\rho(z)\cdot z)$.
\end{definition}

That the function $\iota(f)$ constructed in this way is indeed invariant follows from the equivariance property~\eqref{eq:DefinitionMovingFrame} of the moving frame $\rho$,
\[
 \iota(f)(g\cdot z)=f(\rho(g\cdot z)g\cdot z)=f\big(\rho(z)g^{-1}g\cdot z\big)=f(\rho(z)\cdot z)=\iota(f)(z),
\]
which boils down to the def\/inition of an invariant function~$I$, i.e.\ $I(g\cdot z)=I(z)$. In practice, a~function~$f(z)$ is invariantized by f\/irst transforming its argument using the transformations from~$G$ and then substituting the moving frame for the group parameters. By def\/inition, an invariant that is def\/ined on the jet space~$J^n$ is a \emph{differential invariant}.

Moving frames can also be constructed on a discrete space. In a f\/inite dif\/ference approximation, derivatives of functions are approximated using a f\/inite set of values of these functions, and all the points needed to approximate the derivatives arising in a system of dif\/ferential equations are the points of the stencil introduced in Section~\ref{sec:ConstructionOfInvariantSchemesDordonitsyn}. Because most of the interesting symmetries of dif\/ferential equations that are broken in standard numerical schemes require the use of non-orthogonal discretization meshes, it is benef\/icial to both regard $x$ and $u$ as the dependent variables and the computational variables $\xi$ as the independent variables as was discussed in the previous Section~\ref{sec:ConstructionOfInvariantSchemesMovingMesh}.

Sampling the tuples from the \emph{extended computational space} $M_{\xi}=\{(\xi,z)\}$ at discrete points, i.e.\ at $(\xi_i,z(\xi_i))=(\xi_i,z_i)$, one can introduce the space
$
  M_\xi^{\diamond n}=\{(w_1,\dots,w_n)\, |\, \xi_i\ne \xi_j \textup{ for all}$ $i\ne j\},
$
where $w_i=(\xi_i,z_i)$, which can be identif\/ied as the joint product space of stencil variables. Because the identif\/ier $\xi_i$ of the point $w_i$ is required to be unique, each element of $M_\xi^{\diamond n}$ only includes distinct grid points in the physical space of equation variables. The dimension of the space $M_\xi^{\diamond n}$ depends on the number of independent and dependent variables in the system of dif\/ferential equations and the desired order of accuracy of the approximated derivatives.

It is possible to carry out the construction of the moving frame on $M_\xi^{\diamond n}$, i.e.\ to def\/ine the moving frame by an equivariant mapping $\rho_\xi^{\diamond n}\colon M_\xi^{\diamond n}\to G$, where $G$ acts on $M_\xi^{\diamond n}$ by the product action, $g\cdot (w_1,\dots,w_n)=(g\cdot w_1,\dots,g\cdot w_n)$. Note that the extension of the group action to the computational variables $\xi$ is trivial, i.e.\ they remain unaf\/fected by $G$, $\tilde \xi = g\cdot\xi=\xi$, see~\cite{bihl12By}. The compatibility between the moving frame $\rho_\xi^{\diamond n}$ and the moving frame $\rho$ on the space $M$ (or an appropriate jet space $J^n$), i.e.\ that $\rho_\xi^{\diamond n}\to\rho$ in the continuous limit is assured provided that the cross-section def\/ining the moving frame $\rho_\xi^{\diamond n}$ in the continuous limit converges to the cross-section def\/ining the moving frame $\rho$. Once the moving frame is constructed on the discrete space $M_\xi^{\diamond n}$ of stencil variables, it can be used to invariantize any numerical scheme expressed in computational coordinates. This will be explicitly shown in Sections~\ref{sec:InvariantDiscretizationHeatEquation} and~\ref{sec:NumericalPropertiesSchemesHeatEquation} where we will construct invariant schemes for the heat equation.

It is essential that the construction of the moving frame on the grid point space is carried out in terms of computational coordinates rather than physical coordinates. This can be illustrated by the following simple example.

\begin{example}
 The Laplace equation $u_{xx}+u_{yy}=0$ is, inter alia, invariant under the one-parameter group of rotations $\mathrm{SO}(2)$, $\tilde x=x\cos\varepsilon-y\sin\varepsilon$, $\tilde y=x\sin\varepsilon+y\cos\varepsilon$. Let us obtain the moving frame~$\rho$ for this group action from the normalization condition $u_x=0$, i.e.\ we determine the moving frame on the f\/irst jet space $J^1(M,2)$, $\rho=\rho(x,y,u,u_x,u_y)$. Prolonging the transformations from $\mathrm{SO}(2)$ to the derivative $u_x$ leads to $\tilde u_{\tilde x}=u_x\cos\varepsilon-u_y\sin\varepsilon$ and thus the moving frame is $\varepsilon=\arctan(u_x/u_y)$.

Let us now f\/ind the product frame from the discrete normalization condition $u^{\rm d}_x=0$. Computing $u_x^{\rm d}$ in the na\"{i}ve way, $u^{\rm d}_x=(u_{i+1}-u_{i-1})/(x_{i+1}-x_{i-1})$, we fail as
\[
 \tilde u^{\rm d}_{\tilde x} = \frac{u_{i+1}-u_{i-1}}{(x_{i+1}-x_{i-1})\cos\varepsilon-(y_{i+1}-y_{i-1})\sin\varepsilon}=0,
\]
cannot be solved for the group parameter $\varepsilon$. On the other hand, setting $u=u(x(\xi^1,\xi^2),y(\xi^1,\xi^2))$ and expressing $u_x^{\rm d}$ in terms of the computational variables $\xi^1$, $\xi^2$, the normalization $u_x^{\rm d}=0$ reads
\[
 \tilde u^{\rm d}_{\tilde x} = \frac{\tilde u_{\xi^1}^{\rm d} \tilde y_{\xi^2}^{\rm d}-\tilde u_{\xi^2}^{\rm d}\tilde y_{\xi^1}^{\rm d}}{\tilde x_{\xi^1}^{\rm d} \tilde y_{\xi^2}^{\rm d}-\tilde x_{\xi^2}^{\rm d}\tilde y_{\xi^1}^{\rm d}}=\frac{u_{\xi^1}^{\rm d} \big(x_{\xi^2}^{\rm d}\sin\varepsilon+y_{\xi^2}^{\rm d}\cos\varepsilon\big)-u_{\xi^2}^{\rm d}\big(x_{\xi^1}^{\rm d}\sin\varepsilon+y_{\xi^1}^{\rm d}\cos\varepsilon\big)}{x_{\xi^1}^{\rm d} y_{\xi^2}^{\rm d}-x_{\xi^2}^{\rm d} y_{\xi^1}^{\rm d}}=0.
\]
This expression can be solved for $\varepsilon$ and it gives
\[
 \varepsilon=\arctan\left(\frac{u_{\xi^1}^{\rm d}y_{\xi^2}^{\rm d}-u_{\xi^2}^{\rm d}x_{\xi^1}^{\rm d}}{u_{\xi^2}^{\rm d}y_{\xi^1}^{\rm d}-u_{\xi^1}^{\rm d}x_{\xi^2}^{\rm d}}\right)= \arctan\left(\frac{u_x^{\rm d}}{u_y^{\rm d}}\right),
\]
which in the continuous limit goes to $\varepsilon=\arctan(u_x/u_y)$ as required.
\end{example}

\section{Lie symmetries of the heat equation}\label{sec:HeatEquation}

The one-dimensional linear heat transport equation is
\begin{gather}\label{eq:HeatEquation}
 u_t - u_{xx}=0,
\end{gather}
where we scaled the thermal dif\/fusivity $\nu$ to $1$, i.e.\ equation~\eqref{eq:HeatEquation} is in non-dimensional form.

The heat equation~\eqref{eq:HeatEquation} admits the following inf\/initesimal generators of one-parameter groups, which generate the maximal Lie invariance algebra~$\mathfrak g$ of equation~\eqref{eq:HeatEquation}:
\begin{gather}\label{eq:MaximalLieInvarianceAlgebraHeatEquation}
\begin{split}
 &\partial_t,\quad \partial_x,\quad u\partial_u,\quad 2t\partial_t+x\partial_x,\quad 2t\partial_x - xu\partial_u,\\
 &4t^2\partial_t+4tx\partial_x-(x^2+2t)u\partial_u,\quad \alpha(t,x)\partial_u,
\end{split}
\end{gather}
where $\alpha$ runs through the set of solutions of equation~\eqref{eq:HeatEquation}, see e.g.~\cite{olve86Ay}. These vector f\/ields generate (i)~time-translations, (ii) space translations, (iii)~scalings in~$u$, (iv)~simultaneous scalings in~$t$ and~$x$, (v)~Galilean boosts, (vi)~inversions and (vii)~the superposition principle symmetry.

In this paper, we will construct invariant numerical schemes for a class of initial value problems of the heat equation using periodic boundary conditions. This class of initial-boundary value problems only admits a subgroup of the symmetry group of the heat equation as inversions are no longer admitted; inversions do not send an initial value problem from the considered class to another initial value problem. The symmetries associated with the f\/irst f\/ive vector f\/ields in~\eqref{eq:MaximalLieInvarianceAlgebraHeatEquation} are compatible with the class of initial-boundary value problems we are interested in, i.e.\ they map the class of initial-boundary value problems for the heat equation under consideration to itself. This re-interpretation of symmetries of dif\/ferential equations without initial and boundary conditions as equivalence transformations for a class of initial-boundary value problems was recently pointed out in~\cite{bihl12By}.

In what follows we will thus focus our attention on constructing numerical schemes that preserve the symmetries generated by the f\/irst f\/ive operators of~\eqref{eq:MaximalLieInvarianceAlgebraHeatEquation}. The associated subalgebra of~$\mathfrak g$ will be denoted by~$\mathfrak g^1$. We do not require to preserve the linearity operator here by construction. At the same time, as will be shown in Section~\ref{sec:NumericalTests} the numerical schemes we propose in this paper preserve the linearity property up to the discretization error expected.

\section{Moving frame and dif\/ferential invariants for the heat equation}\label{sec:DifferentialInvariantsHeatEquation}

We determine the moving frame for the subgroup $G^1$ of transformations associated with the subalgebra $\mathfrak g^1$. Transformations of $G^1$ are of the form
\begin{gather}\label{eq:TransformationOfUsedSubgroup}
 \tilde t=e^{2\varepsilon_4}(t+\varepsilon_1),\qquad \tilde x=e^{\varepsilon_4}(x+\varepsilon_2+2\varepsilon_5t),\qquad \tilde u=e^{\varepsilon_3-\varepsilon_5x-\varepsilon_5^2t}u.
\end{gather}
Because the group action of $G^1$ is not free on $M=\{(t,x,u)\}$ we construct the moving frame on a suitably high-order jet space. In the present case, the group action of $G^1$ becomes free on $J^1=J^1(M,2)$. Thus, it is necessary to extend the transformations~\eqref{eq:TransformationOfUsedSubgroup} to derivatives of $u$ with respect to $t$ and $x$.

Using the chain rule we can compute the transformed operators of total dif\/ferentiation, which read as
\[
 \mathrm{D}_{\tilde t}=e^{-2\varepsilon_4}(\mathrm{D}_t-2\varepsilon_5\mathrm{D}_x),\qquad \mathrm{D}_{\tilde x} = e^{-\varepsilon_4}\mathrm{D}_x,
\]
where $\mathrm{D}_x=\partial_x+u_x\partial_u+u_{tx}\partial_{u_t}+u_{xx}\partial_{u_x}+\cdots$ and $\mathrm{D}_t=\partial_t+u_t\partial_u+u_{tt}\partial_{u_t}+u_{tx}\partial_{u_x}+\cdots$ denote the usual operators of total dif\/ferentiation. With the transformed total dif\/ferentiation operators at hand it is possible to compute the transformed partial derivatives of $u$ with respect to $t$ and~$x$. The transformation rules for the lowest order derivatives are
\begin{gather*}
 \tilde u_{\tilde t}=e^{-2\varepsilon_4+\varepsilon_3-\varepsilon_5x-\varepsilon_5^2t}\big(u_t-2\varepsilon_5u_x+\varepsilon_5^2u\big),\qquad \tilde u_{\tilde x} = e^{-\varepsilon_4+\varepsilon_3-\varepsilon_5x-\varepsilon_5^2t}(u_x-\varepsilon_5u),\\
 \tilde u_{\tilde x\tilde x}=e^{-2\varepsilon_4+\varepsilon_3-\varepsilon_5x-\varepsilon_5^2t}\big(u_{xx}-2\varepsilon_5u_x+\varepsilon_5^2u\big).
\end{gather*}
In fact, for the construction of the moving frame already the knowledge of the f\/irst order derivatives is suf\/f\/icient.

We compute the moving frame for the f\/ive-parameter group of transformations of the form~\eqref{eq:TransformationOfUsedSubgroup} using the following normalization conditions which determine a valid cross-section to the group orbits of the prolonged action of~$G^1$ on~$J^1$,
\begin{gather}\label{eq:NormalizationConditions}
 t=0,\qquad x=0,\qquad u=1,\qquad u_t=1,\qquad u_x=0.
\end{gather}
The moving frame is computed by taking the transformed form of the normalization conditions,
$\tilde t=0$, $\tilde x=0$, $\tilde u=1$, $\tilde u_{\tilde t}=1$ and $\tilde u_{\tilde x}=0$ and by solving the resulting algebraic system for the group parameters $\varepsilon_1,\dots,\varepsilon_5$. The result of this computation is the following moving frame~$g=\rho(z^{(1)})$,
\begin{gather}\label{eq:MovingFrame}
\begin{split}
 &\varepsilon_1=-t,\qquad \varepsilon_2=-\left(x+2t\frac{u_x}{u}\right),\qquad \varepsilon_3=-\left(\ln u-x\frac{u_x}{u}-t\frac{u_x^2}{u^2}\right),\\
 &\varepsilon_4=\ln\sqrt{\frac{u_t}{u}-\frac{u_{x}^2}{u^2}},\qquad \varepsilon_5 = \frac{u_x}{u}.
\end{split}
\end{gather}
With the moving frame at hand, we can invariantize any of the partial derivatives of~$u$ with respect to~$t$ and~$x$ and thus obtain a complete set of dif\/ferential invariants for the subgroup $G^1$ of the maximal Lie invariance group of the heat equation. As an example, invariantizing the derivative $u_{xx}$, i.e.\ computing $\iota(u_{xx})$ as $(g\cdot u_{xx})|_{g=\rho(z^{(1)})}$ we produce the dif\/ferential invariant
\[
 \iota(u_{xx})=\frac{uu_{xx}^2-u_{x}^2}{uu_t - u_x^2}.
\]
Invariantizing the heat equation, i.e.\ computing $\iota(u_t-u_{xx})=0$ and recalling that $\iota(u_t)=1$, we obtain
\[
 \frac{u(u_t-u_{xx})}{uu_t-u_x^2}=0,
\]
which yields the original heat equation expressed in terms of dif\/ferential invariants. This re-expression of a dif\/ferential equation using the dif\/ferential invariants of its symmetry group is known as the \emph{replacement theorem}~\cite{cheh08Ay}.

\section{Invariant discretization of the heat equation}\label{sec:InvariantDiscretizationHeatEquation}

The invariant discretization of equation~\eqref{eq:HeatEquation} cannot be done on a f\/ixed, uniform grid. To see this, let us check the transformation behavior of the grid equation
$
 x_i^{n+1}-x_i^n=0,
$
which is the def\/inition of a stationary grid, under the transformations~\eqref{eq:TransformationOfUsedSubgroup}. This yields
\[
 \tilde x_i^{n+1}-\tilde x_i^n=e^{\varepsilon_4}\big(x_i^{n+1}-x_i^n+2\varepsilon_5\big(t^{n+1}-t^n\big)\big),
\]
which is only zero in the case when $\varepsilon_5=0$. Stated in another way, a discretization on a f\/ixed grid can at most preserve the symmetry subgroup of $G$, which is generated by the f\/irst four elements of the maximal Lie invariance algebra $\mathfrak g$ of the heat equation~\eqref{eq:MaximalLieInvarianceAlgebraHeatEquation}.

Thus, the discretization of~\eqref{eq:HeatEquation} preserving~$G^1$ will require the use of moving grids. For this reason it is convenient to express~\eqref{eq:HeatEquation} in terms of computational coordinates initially, i.e.\ we set $u(\tau,\xi)=u(\tau,x(\tau,\xi))$, where $\xi$ is the single spatial computational variable and~$\tau=t$. The heat equation in this set of coordinates reads
\begin{gather}\label{eq:HeatEquationComputationalVariables}
 u_\tau - x_\tau\frac{u_\xi}{x_\xi} - \frac{1}{x_\xi^2}\left(u_{\xi\xi} - \frac{x_{\xi\xi}}{x_\xi}u_\xi\right)=0.
\end{gather}

So as to f\/ind the invariant discretization of the heat equation in the form~\eqref{eq:HeatEquationComputationalVariables}, we determine the moving frame in the space of stencil variables $M_\xi^{\diamond4}$ using the discrete analogs of the normalization conditions~\eqref{eq:NormalizationConditions} expressed in terms of computational coordinates.

\begin{figure}[htdp]
  \centering
  \includegraphics[scale=0.75]{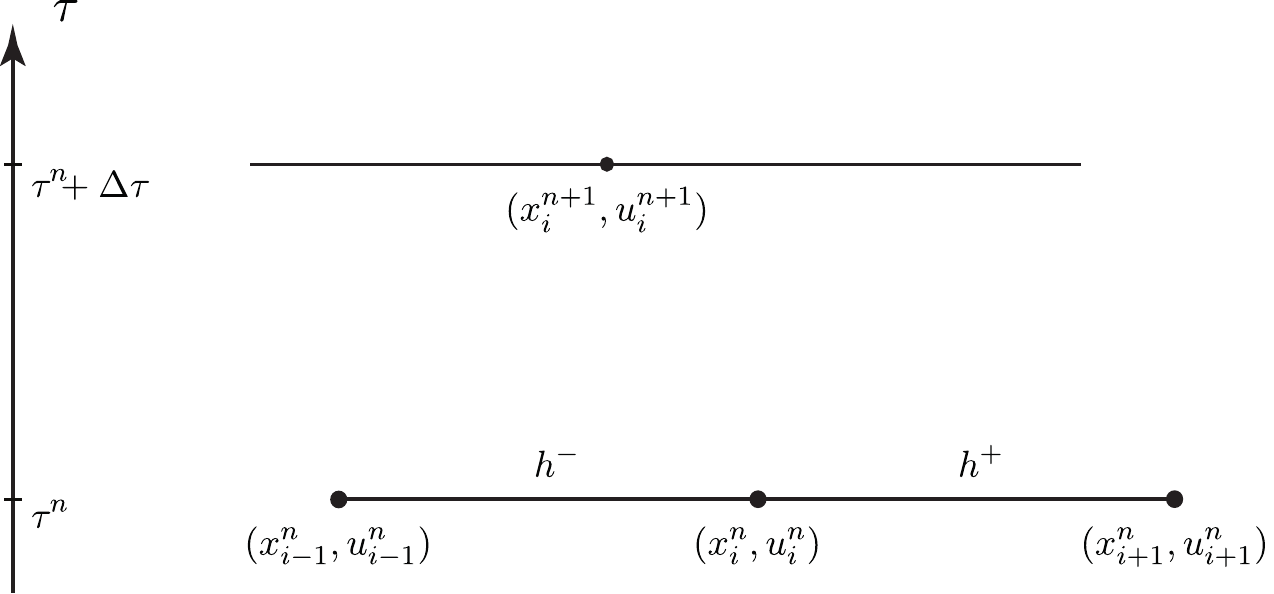}\\
  \caption{Stencil for an invariant discretization scheme of the heat equation.}
\label{fig:StencilHeatEquation}
\end{figure}

For the sake of convenience we introduce the notation $h^+ = x_{i+1}^n-x_i^n$, $h^-=x_i^n-x_{i-1}^n$, $\Delta \tau = \tau^{n+1}-\tau^n$. The discretization stencil we use is depicted in Fig.~\ref{fig:StencilHeatEquation}.

The appropriate normalization conditions for a compatible moving frame $\rho_\xi^{\diamond4}$ are
\begin{gather}\label{eq:NormalizationConditionsDiscrete}
 \tau^n=0,\qquad x_i^n=0,\qquad u_i^n=1,\qquad u_t^{\rm d}=1,\qquad u_x^{\rm d}=0,
\end{gather}
where
\[
 u_t^{\rm d} = \frac{u_i^{n+1}-u_i^n}{\Delta \tau} - x_\tau^{\rm d}\frac{u_{i+1}^n-u_{i-1}^n}{h^++h^-},\qquad u_x^{\rm d} = \frac{u_{i+1}^n-u_{i-1}^n}{h^++h^-}
\]
are the discretizations of the f\/irst time and space derivatives expressed in computational coordinates and $x_\tau^{\rm d}=(x_i^{n+1}-x_i^n)/\Delta\tau$ is the discrete grid velocity. Replacing the single equations in the above normalization conditions by their respective transformed expressions and solving the resulting algebraic system for the group parameters we obtain the following moving frame on the space of stencil variables~$M_\xi^{\diamond4}$,
\begin{gather}\label{eq:DiscreteMovingFrame}
\begin{split}
 &\varepsilon_1=-\tau^n,\qquad \varepsilon_2=-\big(x_i^n+2\tau^n(\ln u)^{\rm d}_x\big),\qquad \varepsilon_3=-\left(\ln u_i^n - x_i^n(\ln u)^{\rm d}_x - \tau^n\big((\ln u)^{\rm d}_x\big)^2\right), \\
 &\varepsilon_4=\frac12\ln\left(\frac{\exp\left({-}\Delta\tau\big(x_\tau^{\rm d}(\ln u)^{\rm d}_x+\big((\ln u)^{\rm d}_x\big)^2\big)\right)u_i^{n+1}-u_i^n}{u_i^n\Delta\tau}\right),\qquad\varepsilon_5=(\ln u)^{\rm d}_x,
\end{split}\hspace{-10mm}
\end{gather}
where we introduced $(\ln u)^{\rm d}_x=(\ln u_{i+1}^n-\ln u_{i-1}^n)/(h^++h^-)$. This moving frame is compatible with the moving frame~\eqref{eq:MovingFrame} in that it converges to~\eqref{eq:MovingFrame} in the continuous limit $\Delta\xi\to0$ and $\Delta\tau\to0$ upon using
\begin{gather}\label{eq:ExpansionOfStencilVariables}
\begin{split}
&h^+ = x_\xi\Delta \xi + x_{\xi\xi}(\Delta \xi)^2/2+O\big(\Delta \xi^3\big),\qquad h^- = x_\xi\Delta \xi - x_{\xi\xi}(\Delta \xi)^2/2+O\big(\Delta \xi^3\big),\\ &x_i^{n+1}=x_i^n+x_\tau\Delta\tau+O\big(\Delta\tau^2\big),\qquad u_{i+1}^n = u_i^n + u_\xi\Delta \xi+ u_{\xi\xi}(\Delta \xi)^2/2 + O\big(\Delta \xi^3\big), \\
&u_{i-1}^n = u_i^n - u_\xi\Delta \xi+ u_{\xi\xi}(\Delta \xi)^2/2 - O\big(\Delta \xi^3\big),\qquad u_i^{n+1} = u_i^n + u_\tau \Delta \tau+O\big(\Delta \tau^2\big).
\end{split}
\end{gather}

The moving frame~\eqref{eq:DiscreteMovingFrame} can now be used to invariantize any non-invariant f\/inite dif\/ference discretization of~\eqref{eq:HeatEquationComputationalVariables} on $M_\xi^{\diamond4}$. To illustrate this, we invariantize the standard FTCS (forward in time centered in space) scheme
\[
 u_t^{\rm d}- \frac{4}{(h^++h^-)^2}\big(u_{i+1}^n+u_{i-1}^n-2u_i^n-(h^+-h^-)u_x^{\rm d}\big)=0.
\]
This is done by f\/irst replacing all terms by their respective transformed expressions and substituting the moving frame~\eqref{eq:DiscreteMovingFrame} for the arising group parameters. The result of this procedure is the invariant scheme
\begin{gather}\label{eq:InvariantSchemeHeatEquation}
\begin{split}
 &S=\frac{\exp\left(-\Delta \tau\left(x_\tau^{\rm d} (\ln u)_x^{\rm d}+((\ln u)_x^{\rm d})^2\right)\right)u_i^{n+1}-u_i^n}{\Delta \tau}\\
 &\hphantom{S=}{}-\frac{4\left(u_{i+1}^n\left(\dfrac{u_{i+1}^n}{u_{i-1}^n}\right)^{-h^+/(h^++h^-)}+ u_{i-1}^n\left(\dfrac{u_{i+1}^n}{u_{i-1}^n}\right)^{h^-/(h^++h^-)}-2u_i^n\right)}{(h^++h^-)^2}=0.
\end{split}
\end{gather}
Again, it can be checked that the above scheme \eqref{eq:InvariantSchemeHeatEquation} indeed converges to \eqref{eq:HeatEquationComputationalVariables} in the limit of $\Delta \xi\to0$ and $\Delta \tau\to0$. This will be shown explicitly in Section~\ref{sec:NumericalPropertiesSchemesHeatEquation}, where we establish the order of approximation of~\eqref{eq:InvariantSchemeHeatEquation}.

So as to complete the scheme~\eqref{eq:InvariantSchemeHeatEquation} it is necessary to determine $x_i^{n+1}$, which is the ingredient missing in~\eqref{eq:InvariantSchemeHeatEquation}. There are dif\/ferent ways to determine a grid equation, such as using the equidistribution principle as outlined in Section~\ref{sec:ConstructionOfInvariantSchemesMovingMesh}. The problem with this strategy in the present case is that while it might be benef\/icial from the numerical point of view, it might not be easy to obtain an invariant discretization of this principle which does not lead to a fully coupled equation--grid system. In other words, it can happen that the grid equation includes values of $u$ at both $t^n$ and $t^{n+1}$. While this coupling is not a problem in the one-dimensional case, it can lead to a severe restriction of the applicability for multi-dimensional equations as solving the coupled equation--grid system might then be too expensive.

A $G^1$-invariant grid equation that circumvents the aforementioned coupling problem can be derived from the invariantization of $x^{n+1}_i$. This invariantization yields
\[
 \iota\big(x^{n+1}_i\big)=e^{\varepsilon_4}\left(x_i^{n+1} - x_i^n + \frac{2\Delta\tau}{h^++h^-}\big(\ln u_{i-1}^n-\ln u_{i+1}^n\big)\right),
\]
where we did not explicitly substitute the frame value for $\varepsilon_4$. An appropriate grid is then given through $\iota(x^{n+1}_i)=0$, or
\begin{gather}\label{eq:InvariantGridEquation}
 M=x_i^{n+1} - x_i^n + \frac{2\Delta\tau}{h^++h^-}\big(\ln u_{i-1}^n-\ln u_{i+1}^n\big)=0.
\end{gather}

This grid equation is quite similar to the grid equation
\begin{gather}\label{eq:InvariantGridEquationDorodnitsyn}
 x_i^{n+1} - x_i^n + \frac{2\Delta\tau}{h^++h^-}\left(\frac{h^+}{h^-}\ln\left(\frac{u_{i-1}^n}{u_i^n}\right)-\frac{h^-}{h^+}\ln\left(\frac{u_{i+1}^n}{u_i^n}\right)\right)=0,
\end{gather}
which was found in~\cite{baki97Ay,doro03Ay} using the method of dif\/ference invariants. This last grid~\eqref{eq:InvariantGridEquationDorodnitsyn} is not only invariant under the subgroup $G^1$ but under the whole maximal Lie invariance group $G$ of the heat equation. In the continuous limit, both equation~\eqref{eq:InvariantGridEquation} and~\eqref{eq:InvariantGridEquationDorodnitsyn} converge to
\[
 x_\tau = -\frac{2}{x_\xi}(\ln u)_\xi.
\]

We have tested all our numerical schemes with both~\eqref{eq:InvariantGridEquation} and~\eqref{eq:InvariantGridEquationDorodnitsyn} and found that the resulting schemes give asymptotically the same numerical results. In fact, as in the evolution--projection strategy that will be introduced in Section~\ref{sec:InvariantInterpolationSchemes} we have $h^+=h^-=h$, equation~\eqref{eq:InvariantGridEquation} and~\eqref{eq:InvariantGridEquationDorodnitsyn} coincide.

\begin{remark}\label{rem:CheckingInvarianceOfScheme}
While the invariantization algorithm guarantees that the scheme~\eqref{eq:InvariantSchemeHeatEquation} is indeed invariant under the subgroup $G^1$ of the maximal Lie invariance group of the heat equation, the invariance can be checked in a straightforward fashion using the inf\/initesimal invariance criterion as invoked in the Dorodnitsyn method discussed in Section~\ref{sec:ConstructionOfInvariantSchemesDordonitsyn}. Let us recall that this criterion states that an invariant $I$ of a group action is annihilated by the associated inf\/initesimal generators, i.e.\ $\mathrm{v}(I)=0$ for all $\mathrm{v}\in\mathfrak g$. Because in the present case, the invariants are def\/ined on the stencil space with coordinates $\tau^n$, $\Delta \tau$, $x_i^n$, $x_{i+1}^n$, $x_{i-1}^n$, $x_i^{n+1}$, $u_i^n$, $u_{i+1}^n$, $u_{i-1}^n$ and $u_i^{n+1}$, we have to prolong the operators of~$\mathfrak g$ accordingly. The prolongations of the f\/irst f\/ive operators of~\eqref{eq:MaximalLieInvarianceAlgebraHeatEquation} to the variables of the stencil are
\begin{gather*}
  \partial_{\tau^n},\qquad \partial_{x_i^n}+\partial_{x_{i+1}^n}+\partial_{x_{i-1}^n}+\partial_{x_i^{n+1}},\qquad u_i^n\partial_{u_i^n}+u_{i+1}^n\partial_{u_{i+1}^n}+u_{i-1}^n\partial_{u_{i-1}^n}+u_i^{n+1}\partial_{u_i^{n+1}},\\
  2\tau^n\partial_{\tau^n}+2\Delta\tau\partial_{\Delta\tau}+x_i^n\partial_{x_i^n}+x_{i+1}^n\partial_{x_{i+1}^n}+ x_{i-1}^n\partial_{x_{i-1}^n}+x_i^{n+1}\partial_{x_i^{n+1}},\\
 2\tau^n(\partial_{x_i^n}+\partial_{x_{i+1}^n}+
  \partial_{x_{i-1}^n})+2(\tau^n+\Delta\tau)\partial_{x_i^{n+1}}-x_i^nu_i^n\partial_{u_i^n}-x_{i+1}^nu_{i+1}^n\partial_{u_{i+1}^n}\\
  \qquad{} -x_{i-1}^nu_{i-1}^n\partial_{u_{i-1}^n}-x_i^{n+1}u_i^{n+1}\partial_{u_i^{n+1}},
\end{gather*}
see~\cite{doro11Ay,levi06Ay} for more details. It can be checked that $\mathrm{pr}\,\mathrm{v}(S)=0$ and $\mathrm{pr}\,\mathrm{v}(M)=0$ hold on $S=0$ and $M=0$ for all the prolonged inf\/initesimal generators and thus $S=0$ is a proper invariant numerical scheme and $M=0$ an invariant grid equation.
\end{remark}

\begin{remark}
 The heat equation is a linear partial dif\/ferential equation in two independent variables. One might thus consider to set up a grid equation not only for spatial but also for temporal adaptation. The reason why we refrain from spatial-temporal adaptation here is that the symmetry group $G$ of the heat equation is compatible with f\/lat time layers, i.e. $t^{n}_{i+1}-t^n_i=0$ is a~$G$-invariant equation. Improving the invariant numerical scheme constructed above using temporal adaptation would thus not allow one a fair comparison against the original non-invariant FTCS scheme for the heat equation. Moreover, f\/lat time layers are well-agreed with the physics of the heat transfer problem, which af\/fects all points of the domain simultaneously.
\end{remark}

\section{Numerical properties of the invariant scheme}\label{sec:NumericalPropertiesSchemesHeatEquation}

In this section we investigate the numerical properties of the scheme~\eqref{eq:InvariantSchemeHeatEquation} and related schemes. We start our consideration with the estimation of the local truncation error of the scheme. The study of this question is relevant because so far little is known about the relation between the order of a non-invariant scheme and its invariantized counterpart.

The discretization of the heat equation in computational coordinates~\eqref{eq:HeatEquationComputationalVariables} can be formally represented as
\begin{gather}\label{eq:HeatEquationComputationalVariablesDiscretization}
 u_\tau^{\rm d} - x_\tau^{\rm d}\frac{u_\xi^{\rm d}}{x_\xi^{\rm d}} - \frac{1}{(x_\xi^{\rm d})^2}\left(u_{\xi\xi}^{\rm d} - \frac{x_{\xi\xi}^{\rm d}}{x_\xi^{\rm d}}u_\xi^{\rm d}\right)=0,
\end{gather}
where in the present case we assume that derivatives are approximated with the aid of a standard FTCS scheme. More general schemes will be considered after the order of the invariantized FTCS scheme is established.

\begin{theorem}\label{thm:TheoremOnFirstOrderInvariantScheme}
 The order of the invariant scheme~\eqref{eq:InvariantSchemeHeatEquation} is the same as the order of the scheme~\eqref{eq:HeatEquationComputationalVariablesDiscretization}, namely first order in time and second order in space, provided that an Euler forward step and second order centered differences are used to approximate the time and space derivatives arising in both the differential equation~\eqref{eq:HeatEquationComputationalVariablesDiscretization} and the normalization conditions~\eqref{eq:NormalizationConditionsDiscrete}.
\end{theorem}

\begin{proof}
Invariantizing the scheme~\eqref{eq:HeatEquationComputationalVariablesDiscretization} using the normalization conditions~\eqref{eq:NormalizationConditionsDiscrete} leads to
\begin{gather}\label{eq:GeneralInvariantizedFormOfScheme}
 1 - \frac{4}{\iota\big(\big(x_\xi^{\rm d}\big)^2\big)}\iota\big(u_{\xi\xi}^{\rm d}\big)=0,
\end{gather}
where $\iota(f)(z)$ denotes the invariantization of the function $f(z)$. By def\/inition, invariantization of a function $f(z)$ means to transform the argument $z$ and plug in the moving frame for the group parameters. In the present case, the transformed form of~\eqref{eq:GeneralInvariantizedFormOfScheme} can be written as
\[
 1 - \frac{4e^{\varepsilon_3-\varepsilon_5x-\varepsilon_5^2t-2\varepsilon_4}}{\big(x_\xi^{\rm d}\big)^2}\big(e^{-\varepsilon_5h^+}u_{i+1}^n + e^{\varepsilon_5h^-}u_{i-1}^n -2u_i^n\big)=0.
\]
Using the normalization condition $u_i^n=1$ we obtain that
\[
 \tilde u_i^n = 1 = e^{\varepsilon_3-\varepsilon_5x-\varepsilon_5^2t}u_i^n
\]
and thus the last expression can be recast as
\begin{gather}\label{eq:TransformedScheme}
e^{2\varepsilon_4}u_i^n-\frac{4}{(h^++h^-)^2}\big(e^{-\varepsilon_5h^+}u_{i+1}^n + e^{\varepsilon_5h^-}u_{i-1}^n -2u_i^n\big)=0.
\end{gather}
Let us now determine the local discretization error in the parameter $\varepsilon_5$. The respective moving frame component is
\[
 \varepsilon_5=\frac{\ln u_{i+1}^n-\ln u_{i-1}^n}{h^++h^-},
\]
which upon using~\eqref{eq:ExpansionOfStencilVariables} expands~to
\begin{gather}\label{eq:MovingFrameComponent5}
 \varepsilon_5 = \frac{1}{x_\xi}\frac{u_\xi}{u_i^n}+O\big(\Delta \xi^2\big).
\end{gather}
Substituting $\varepsilon_5$ into the second term of equation~\eqref{eq:TransformedScheme} and expanding the exponential functions in the same term into Taylor series, we obtain after some rearranging
\begin{gather*}
 e^{2\varepsilon_4}u_i^n - \frac{1}{x_\xi^2\Delta\xi^2+O\big(\Delta\xi^4\big)}\bigg(u_{i+1}^n + u_{i-1}^n -2u_i^n - \varepsilon_5\bigg(x_\xi\Delta \xi\big(u_{i+1}^n - u_{i-1}^n\big)\\
\qquad{}+\frac12x_{\xi\xi}\Delta\xi^2\big(u_{i+1}^n+u_{i-1}^n\big)\bigg)+
\frac12\varepsilon_2^2x_\xi^2\Delta\xi^2\big(u_{i+1}^n+u_{i-1}^n\big)+O\big(\Delta\xi^4\big)\bigg)=0.
\end{gather*}
This can be further simplif\/ied to
\begin{gather}\label{eq:SpatialDiscretizationScheme}
 e^{2\varepsilon_4}u_i^n - \frac{1}{x_\xi^2}\left(u_{\xi\xi} - \frac{u_\xi^2}{u_i^n} - \frac{x_{\xi\xi}}{x_\xi}u_\xi\right) +O\big(\Delta\xi^2\big)=0.
\end{gather}
It now remains to expand the f\/irst term in equation~\eqref{eq:SpatialDiscretizationScheme}. The moving frame component for $\varepsilon_4$ in~\eqref{eq:DiscreteMovingFrame} can be recast as
\begin{gather*}
 e^{2\varepsilon_4}u_i^n=\frac{\exp\left(-\Delta\tau\left(\dfrac{x_i^{n+1}-x_i^n}{\Delta\tau}\dfrac{\ln u_{i+1}^n-\ln u_{i-1}^n}{h^+-h^-}+\left(\dfrac{\ln u_{i+1}^n-\ln u_{i-1}^n}{h^+-h^-}\right)^2\right)\right)u_i^{n+1}-u_i^n}{\Delta\tau}.
\end{gather*}
Using $x_i^{n+1}=x_i^n+x_\tau\Delta\tau+O(\Delta\tau^2)$ and $u_i^{n+1} = u_i^n + u_\tau \Delta \tau+O(\Delta \tau^2)$ and again expanding the exponential function into a Taylor series, we derive
\[
 e^{2\varepsilon_4}u_i^n= u_\tau - x_\tau\frac{u_\xi}{x_\xi} - \frac{1}{x_\xi^2}\frac{u_\xi^2}{u_i^n} + O\big(\Delta \tau,\Delta\xi^2\big).
\]
Plugging this into equation~\eqref{eq:SpatialDiscretizationScheme} we arrive at
\[
 u_\tau - x_\tau\frac{u_\xi}{x_\xi} - \frac{1}{x_\xi^2}\left(u_{\xi\xi} - \frac{x_{\xi\xi}}{x_\xi}u_\xi\right) + O\big(\Delta \tau,\Delta\xi^2\big)=0,
\]
which completes the proof of the theorem.
\end{proof}

A more general statement is the following one:
\begin{theorem}\label{thm:TheoremOnArbitrarySpatialOrderInvariantSchemes}
The order of spatial discretization of an invariant finite difference scheme for the heat equation in computational variables equals the order $p\in\mathbb{N}$ of the spatial discretization of the associated non-invariant finite difference scheme provided that centered differences of order~$p$ are used to approximate both the derivatives in the heat equation and in the normalization conditions~\eqref{eq:NormalizationConditionsDiscrete}.
\end{theorem}

\begin{proof}
In view of the general form~\eqref{eq:GeneralInvariantizedFormOfScheme} of the invariantization of scheme~\eqref{eq:HeatEquationComputationalVariablesDiscretization}, we study the invariantization of the terms $x_\xi$ and $u_{\xi\xi}$.

The invariantization of $(x_\xi^{\rm d})^2=(x_\xi+O(\Delta \xi^p))^2$ is $\iota((x_\xi^{\rm d})^2)=e^{2\varepsilon_4}(x_\xi^2+O(\Delta \xi^p$)) and is of the same order $p$ if, as required, the moving frame component $\varepsilon_4$ stems from the approximation of $u_\tau^{\rm d}=1$ using $p$th order accuracy and thus only includes approximations of derivatives with that accuracy.

Let us now investigate the invariantization of discretizations of $u_{\xi\xi}$. The general form of a~centered dif\/ference approximation of $u_{\xi\xi}$ of even order~$p$ is
 \[
 u_{\xi\xi} = \frac{1}{\Delta\xi^2}\sum_{j=-p/2}^{p/2}c^2_{p,j}u_j^n+O\big(\Delta\xi^p\big),
 \]
where $c^2_{p,j}=2c^1_{p,j}/j$, $j\in A=\{-p/2,\dots,-1,1,\dots,p/2\}$, $c^2_{p,0}=-2\sum\limits_{i=1}^{p/2}1/i^2$ and
\[
 c^1_{p,j}=\frac{(-1)^{j+1}(p/2)!^2}{j(p/2+j)!(p/2-j)!},\qquad j\in A
\]
and $c^1_{p,0}=0$ are the coef\/f\/icients from the $p$th order approximation of $u_\xi$, i.e.\
\[
 u_{\xi} = \frac{1}{\Delta\xi}\sum_{j=-p/2}^{p/2}c^1_{p,j}u_j^n+O\big(\Delta\xi^p\big).
\]
See~\cite{forn96Ay} for a discussion of the algorithm for f\/inding the weights $c^k_{p,j}$ in higher-order f\/inite dif\/ference approximations of the $k$th derivative of~$u$. The invariantization of $u_{\xi\xi}$ is
\[
 \iota(u_{\xi\xi}) = \frac{1}{\Delta\xi^2}\sum_{j=-p/2}^{p/2}c^2_{p,j} \exp\big(\varepsilon_3-\varepsilon_5x_j^n-\varepsilon_5^2\tau^n\big)u_j^n,
\]
or, upon using the normalization condition $u_i^n=1$,
\begin{gather}\label{eq:TransformedSecondOrderDerivative}
 \iota(u_{\xi\xi}) = \frac{1}{\Delta\xi^2}\frac{1}{u_i^n}\sum_{j=-p/2}^{p/2}c^2_{p,j} \exp(-\varepsilon_5\Delta x_j)u_j^n,
\end{gather}
where we expand
\[
 \Delta x_j=x_j^n-x_i^n=\sum_{k=1}^{\infty}\frac{(j \Delta \xi)^k}{k!}\frac{\partial^kx}{\partial\xi^k},\qquad u_i^n=\sum_{l=0}^{\infty}\frac{(j\Delta \xi)^l}{l!}\frac{\partial^lu}{\partial\xi^l}.
\]
Using the expressions for $\Delta x_j$ and $u_i^n$, the expression~\eqref{eq:TransformedSecondOrderDerivative} can be expanded and rearranged in powers of $j\Delta\xi$ in the form
\[
 \iota(u_{\xi\xi})=\frac{1}{\Delta\xi^2}\frac{1}{u_i^n}\sum_{j=-p/2}^{p/2}\sum_{k=0}^\infty (-1)^k c^2_{p,j} A_k(j\Delta\xi)^k,
\]
where
\[
 A_2=\frac12\left(u_{\xi\xi}-\varepsilon_5\big(2x_\xi u_\xi+x_{\xi\xi}u_i^n\big)+\varepsilon_5^2x_\xi^2u_i^n\right).
\]
The expressions for $A_k$, $k\ne2$, are not required subsequently. The proof is completed upon substituting for~$\varepsilon_5$ the corresponding moving frame component (which is of order~$p$ if the normalization $u_x^{\rm d}=0$ is approximated with~$p$th order accuracy) and by noting that
\[
 \sum_{j=-p/2}^{p/2}c^2_{p,j}j^k= \begin{cases}0 & \textup{for } k\in\{0,1,3,\dots p+2,2n\},\quad n\in\mathbb{N}, \\ 2 &  \textup{for } k=2, \\
 c_k\ne0 & \textup{else}, \end{cases}.
\]
where the precise values of the constants $c_k$ follow from evaluating the respective sums.
\end{proof}

The scheme~\eqref{eq:InvariantSchemeHeatEquation} is only of f\/irst order in time $\tau=t$. To construct a scheme that is second order in time, we can start with a non-invariant scheme~\eqref{eq:HeatEquationComputationalVariablesDiscretization} and discretize the time derivative~$u_\tau^{\rm d}$ with second order accuracy, i.e.\ we set $u_\tau^{\rm d}=(u_i^{n+1} -u_i^{n-1})/(2\Delta\tau)$, where $u_i^{n-1}$ is the value of $u$ at the previous time step $\tau^{n-1}=\tau^n-\Delta\tau$. It is now necessary to check whether invariantizing this leapfrog discretization leads to an invariant scheme that is also second order in time.

\begin{theorem}\label{thm:TheoremOnInvariantLeapfrogScheme}
 Invariantization of the scheme~\eqref{eq:HeatEquationComputationalVariablesDiscretization} in which a leapfrog step and second order centered differences are used to approximate the time and space derivatives, leads to an invariant scheme that is both second order in time and space provided that the normalization conditions~\eqref{eq:NormalizationConditionsDiscrete} are approximated using discretizations that are of second order.
\end{theorem}

\begin{proof}
 To prove this theorem it is suf\/f\/icient to establish the order of the f\/irst term in equation~\eqref{eq:SpatialDiscretizationScheme}. We proceed in an analog manner as in the proof of Theorem~\ref{thm:TheoremOnFirstOrderInvariantScheme}, i.e.\ we discretize the normalization condition $u_\tau^{\rm d}=1$ but now with second order accuracy. This yields
 \[
 \tilde u_{\tilde\tau}^{\rm d}=\frac{e^{\varepsilon_3-\varepsilon_5x_i^n-\varepsilon_5^2\tau^n}}{2e^{2\varepsilon_4}\Delta\tau}\big(e^{-\varepsilon_5 x_\tau\Delta\tau-\varepsilon_5^2\Delta\tau}u_i^{n+1} - e^{\varepsilon_5 x_\tau\Delta\tau+\varepsilon_5^2\Delta\tau}u_i^{n-1}\big) =1.
 \]
 Using the normalization condition $u_i^n=1$ as before and expanding the exponential functions we derive
 \[
  e^{2\varepsilon_4}u_i^n=\frac{1}{2\Delta\tau}\left(u_i^{n+1}-u_i^{n-1}-\Delta\tau\big(\varepsilon_5 x_\tau+\varepsilon_5^2\big)\big(u_i^{n+1}+u_i^{n-1}\big)\right)+O\big(\Delta\tau^2\big)
 \]
 and upon noting that $u_i^{n+1}+u_i^{n-1} = 2u_i^n+O(\Delta\tau^2)$ we obtain
 \[
 e^{2\varepsilon_4}u_i^n=u_\tau- x_\tau\frac{u_\xi}{x_\xi} - \frac{1}{x_\xi^2}\frac{u_\xi^2}{u_i^n} + O\big(\Delta \tau^2,\Delta \xi^2\big),
 \]
where we have substituted the expression~\eqref{eq:MovingFrameComponent5} for~$\varepsilon_5$. Plugging this result into equation~\eqref{eq:SpatialDiscretizationScheme} completes the proof of the theorem.
\end{proof}

The actual form of the resulting invariant leapfrog scheme is
\begin{gather*}%\label{eq:InvariantSchemeHeatEquationLeapfrog}
%\begin{split}
 \frac{\exp\left(-\Delta \tau\left(\hat x_\tau^{\rm d} (\ln u)_x^{\rm d}+((\ln u)_x^{\rm d})^2\right)\right)u_i^{n+1}-\exp\left(\Delta \tau\left(\check x_\tau^{\rm d} (\ln u)_x^{\rm d}+((\ln u)_x^{\rm d})^2\right)\right)u_i^{n-1}}{2\Delta \tau} \\
\qquad{}-\frac{4\left(u_{i+1}^n\left(\dfrac{u_{i+1}^n}{u_{i-1}^n}\right)^{-h^+/(h^++h^-)}+
 u_{i-1}^n\left(\dfrac{u_{i+1}^n}{u_{i-1}^n}\right)^{h^-/(h^++h^-)}-2u_i^n\right)}{(h^++h^-)^2}=0,
%\end{split}
\end{gather*}
where $\hat x_\tau^{\rm d}=(x_i^{n+1}-x_i^n)/\Delta\tau$ and $\check x_\tau^{\rm d}=(x_i^n-x_i^{n-1})/\Delta\tau$.

Higher order in time schemes can be constructed upon invariantizing multi-stage schemes. Combining this result with the result established in Theorem~\ref{thm:TheoremOnArbitrarySpatialOrderInvariantSchemes} we have found the following:

\begin{corollary}
 Invariantizing a non-invariant finite difference scheme for the heat equation in computational coordinates preserves the spatial and temporal order of the initial non-invariant finite difference scheme provided that centered differences are used and the normalization conditions for the moving frame are discretized with the same order as the respective derivatives in the non-invariant finite difference scheme.
\end{corollary}

\section{Invariant interpolation schemes}\label{sec:InvariantInterpolationSchemes}

A common property of invariant numerical schemes for evolution equations possessing a non\-trivial maximal Lie invariance group is that it is not possible to use a f\/ixed, orthogonal discretization mesh. The continuous evolution of the mesh, if not handled properly, can lead to several undesirable properties, such as an overly strong concentration of grid points in certain regions and therefore too poor a resolution in other parts of the integration domain. The problem gets worse in the multi-dimensional case where mesh tangling or strongly skewed meshes can occur. But even if the mesh movement can be managed in an optimal way there are various physical problems for which continuously adapting grids pose a severe challenge. An example for this are practically all models that are in operational use in weather and climate prediction. These models employ sophisticated data assimilation strategies and are coupled to subgrid-scale parameterizations that aim to mimic the ef\/fects of unresolved processes on the grid scale variables. Attempting to make use of data assimilation or parameterization schemes on moving meshes is not only a technical problem that would cause a signif\/icant computational overhead compared to standard schemes but also a conceptual challenge for it is unclear on how to design parameterization schemes that can operate on grids with varying resolution. In order to promote the ideas of invariant numerical discretization schemes beyond their application to simple evolution equations it is thus instructive to study possible ways of overcoming the limitations imposed by the requirement of using moving meshes.

One straightforward idea is to use invariant schemes on f\/ixed (i.e.\ non-invariant grids). As was shown in~\cite{rebe11Ay} this can lead to improved numerical solutions compared to non-invariant integrators, while still being excelled by the results that can be obtained using completely invariant schemes. On the other hand, if moving (invariant) meshes are not tractable for a~particular class of problems, preserving the invariance of a system of dif\/ferential equations at least for the discretization of the system itself might be a possible trade-of\/f to take.

Another idea is to use an \textit{evolution--projection strategy}, which will be proposed in the following. This concept relies on using the invariant scheme for the system of dif\/ferential equations together with the invariant mesh equations for a single time step followed by the projection of the numerical solution back to the regular mesh. A similar strategy has proven successful in semi-Lagrangian time integration schemes~\cite{stan91Ay}.

In the present case, the projection step can be practically realized by using \textit{interpolation}. Obviously, any standard interpolation scheme can be used to map the numerical solution $u_i^{n+1}$ def\/ined at $x_i^{n+1}$ to the uniformly spaced $\xi$-grid. This, however, can break the invariance of the numerical scheme as a whole and so the question arises whether it is possible to accomplish the interpolation step in a symmetry-preserving fashion.

In the following we discuss two possible ways of formulating \textit{invariant interpolation schemes}, both of which can be used for f\/inding interpolations that allow the re-mapping of the numerical solution on a moving mesh to a f\/ixed, Cartesian, equally-spaced grid. These ways are the \textit{invariantization of non-invariant interpolation schemes} with moving frames and the construction of \textit{interpolations using difference invariants}.

\medskip

\textbf{Invariantization of interpolation schemes.} The moving frame constructed in the course of invariantizing a f\/inite dif\/ference scheme can also be used to invariantize a certain interpolation method. We exemplify this idea by invariantizing the formula for linear interpolation,
\[
 u_i^{n+1}(y)=u_i^{n+1}+\big(y-x_i^{n+1}\big)\frac{u_{i+1}^{n+1}-u_i^{n+1}}{x_{i+1}^{n+1}-x_i^{n+1}},
\]
where $y\in[x_i^{n+1},x_{i+1}^{n+1}]$. The invariantization of this expression using a moving frame associated with $G^1$ yields the invariant interpolation formula
\begin{gather}\label{eq:InvariantLinearInterpolation}
\begin{split}
&  u_i^{n+1}(y)=U_i^{n+1}+\big(y-x_i^{n+1}\big)\frac{U_{i+1}^{n+1}-U_i^{n+1}}{x_{i+1}^{n+1}-x_i^{n+1}},\\
 &U_i^{n+1}=\exp \big((\ln \hat u_{\hat x})^{\rm d}\big(y-x_i^{n+1}\big)\big)u_i^{n+1}.
 \end{split}
\end{gather}
Note that we have used a slightly dif\/ferent moving frame for the invarianization as we have used for invariantizing the f\/inite dif\/ference scheme for the heat equation. Specif\/ically, this moving frame is constructed by replacing the normalization condition $u_{x}^{\rm d}=0$ with $\hat u_{\hat x}^{\rm d}=0$. The reason for this is that the moving frame used earlier yielded $\varepsilon_5=(\ln u_x)^{\rm d}$, i.e.\ it involves the solutions of~$u_i$ at the time step~$\tau^n$ rather than at~$\tau^n+\Delta\tau$. Irrespectively of what normalization is used, both interpolations are invariant. Setting $y=\xi_i$ in the above interpolation formula yields~$u_i^{n+1}$ on the regular computational grid. Note that the interpolation~\eqref{eq:InvariantLinearInterpolation} is consistent in that $u_i^{n+1}(x_i^{n+1})=u_i^{n+1}$ and $u_i^{n+1}(x_{i+1}^{n+1})=u_{i+1}^{n+1}$.

In a similar manner more sophisticated interpolation schemes can be invariantized. In the following, we will use the invariantization of quadratic interpolation. Usual quadratic interpolation is based on the expression
\begin{gather}\label{eq:QuadraticInterpolationNonInvariant}
u^{n+1}(y) = u_{i-1}^{n+1}L_{i-1}(y)+u_i^{n+1}L_i(y)+u_{i+1}^{n+1}L_{i+1}(y),\qquad L_j(y)=\prod_{k=i-1 \atop k\ne j}^{i+1}\frac{y-x_k^{n+1}}{x_j^{n+1}-x_k^{n+1}},
\end{gather}
where $y\in[x_{i-1}^{n+1},x_{i+1}^{n+1}]$ and $L_j(y)$ are the Lagrangian interpolation polynomials. Invariantizing this formula using the same moving frame as above we get
\begin{gather}\label{eq:QuadraticInterpolationInvariant}
u^{n+1}(y) = U_{i-1}^{n+1}L_{i-1}(y)+U_i^{n+1}L_i(y)+U_{i+1}^{n+1}L_{i+1}(y),
\end{gather}
where $U_i=\exp ((\ln \hat u_{\hat x})^{\rm d}(\hat x-x_i^{n+1}))u_i^{n+1}$, as in the case of the invariant linear interpolation~\eqref{eq:InvariantLinearInterpolation}. Numerical examples using the invariant quadratic interpolation will be given in Section~\ref{sec:NumericalTests}.

\medskip

 \textbf{Interpolation using dif\/ference invariants.} The product frame on the grid point space allows invariantizing the elementary variables~$x_i^n$ and~$u_i^n$, which yields the system of joint invariants. In the continuous limit these invariants take the normalization values chosen for~$x$ and~$u$ to construct the usual moving frame $\rho$~\cite{olve01Ay}. On the other hand, on the discrete space $M^{\diamond n}_\xi$ we only normalize one~$x_i^n$ ($i,n$ f\/ixed) among all the grid points $x_l^k$ and the analog statement is true for the associated values $u_l^k$. This means that the joint invariants $\iota(x_l^k)$ and $\iota (u_l^k)$, $l\ne i,k\ne n$, are nontrivial and can be used to assemble invariant interpolation schemes.

In the present case, while we have normalized $u_i^n=1$ in the course of constructing the moving frame $\rho^{\diamond 4}$, we are free to use the moving frame to invariantized any~$u_l^k$ where $l\ne i, k\ne n$ and this will yield a proper (nontrivial) invariant on the discrete space $M^{\diamond 4}_\xi$. As above, we again recompute the moving frame for $G^1$ by replacing the normalization conditions $u_i^n=1$ and $u_x^{\rm d}=0$ with $u_i^{n+1}=0$ and $\hat u_{\hat x}^{\rm d}=0$, respectively, which yields new expressions for the moving frame components of $\varepsilon_3$ and $\varepsilon_5$ given by
\[
 \varepsilon_3=-\big(\big(\ln u_i^{n+1}-x_i^{n+1}(\ln \hat u_{\hat x})^{\rm d}-\tau^{n+1}\big((\ln u_{\hat x})^{\rm d}\big)\big)^2\big),\qquad \varepsilon_5=(\ln \hat u_{\hat x})^{\rm d}.
\]
Using this modif\/ied moving frame, we then invarianize the variable $u^{n+1}(\xi_i)$, which is the sought value of~$u$ at the point $(\tau^{n+1},\xi_i)$ of the computational domain. This invariantization yields
\[
 \iota\big(u^{n+1}(\xi_i)\big)= \frac{u^{n+1}(\xi_i)}{u_i^{n+1}}\exp\big((\ln \hat u_{\hat x})^{\rm d}\big(x_i^{n+1}-\xi_i\big)\big).
\]
Because in the continuous limit the invariantization of $u^{n+1}(\xi_i)$ must reproduce the normalization condition $u=1$, we restrict the dif\/ference invariant to the manifold $\iota(u^{n+1}(\xi_i))=1$. The invariant interpolation is thus
\begin{gather*}%\label{eq:InvariantInterpolationDifferenceInvariants}
 u^{n+1}(\xi_i)=\exp\big((\ln \hat u_{\hat x})^{\rm d}\big(\xi_i-x_i^{n+1}\big)\big)u_i^{n+1}
\end{gather*}
and it is again consistent as $u^{n+1}(x_i^{n+1})=u_i^{n+1}$. More accurate interpolations could be constructed by combining the invariants $\iota(x_l^k)$ and $\iota(u_l^k)$ in a suitable way.

\medskip

 The advantage of the interpolation methods introduced in this section is that they are invariant under the group $G^1$, i.e.\ using these interpolation formulas to map $u_i^{n+1}$ back to the $\xi$-grid does not break the invariance of the numerical schemes for the heat equation, while still allowing one to use a regular grid. Invariant interpolations thus allow avoiding the complications that moving meshes impose on the applicability of symmetry-preserving f\/inite dif\/ference discretizations.

\section{Numerical verif\/ication}\label{sec:NumericalTests}

In order to verify the accuracy predicted above for the various schemes proposed, we set-up the following problem. On a periodic domain $x\in\left[0,2\pi\right[$, consider
\begin{gather*}
 u_t = u_{xx},\\
 u (x,t=0 ) = \sin(x-1)+2.
\end{gather*}
On a sequence of grids with $N\in \{ 2,4,8,\dots,256 \} $, the number of grid points, we compute the error in the maximum norm between the numerical solution and the exact solution at $t=1$. The time step $\Delta\tau$ is taken as proportional to $h^2$, $h=h^+=h^-$, in all simulations.

In each of the following f\/igures, we plot the reference line corresponding to $\mathcal{O}\big(h^{2}\big)$ dash--dotted, and the $L_{\infty}$ error as black line where,
\[
 \Vert E \Vert _{L_{\infty}}=\underset{x\in [0,2\pi ]}{\max}\left|u (x,1 )-u_{\rm exact} t(x,1 )\right|.
\]

Note that for all approaches described below we expect a second order convergence since the numerical scheme being used is of second order, its invariantization was shown to preserve this order and the quadratic interpolation and its invariantization is also of second order.

\subsection{Invariant scheme without projection}

\begin{figure}[ht!]
\centering
\includegraphics[scale=0.38]{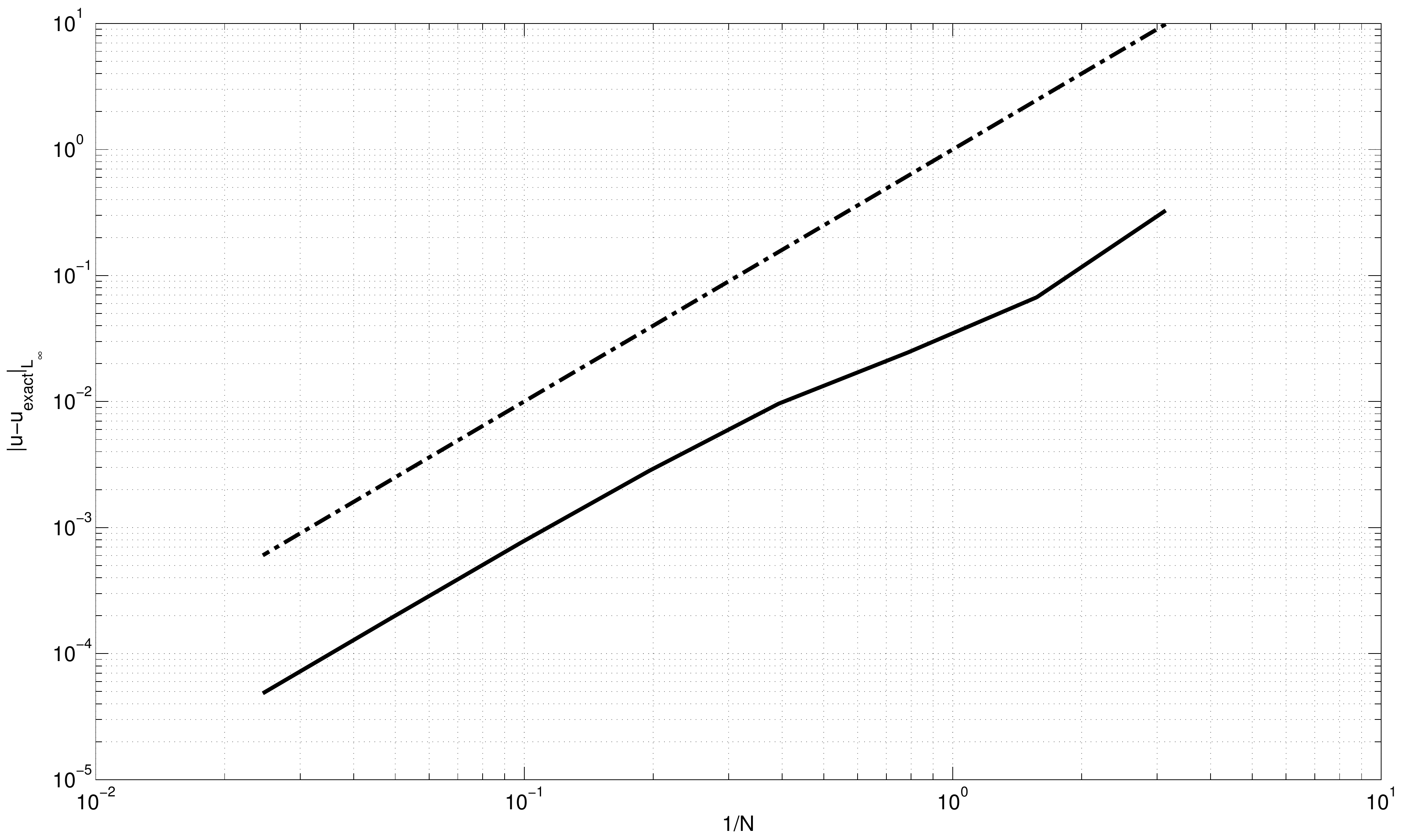}
\caption{Convergence plot for the invariant scheme~\eqref{eq:InvariantSchemeHeatEquation} with invariant grid equation~\eqref{eq:InvariantGridEquation} and without re-mapping.}
\label{fig:convergence_no_remapping}
\end{figure}

In this test run we use the scheme~\eqref{eq:InvariantSchemeHeatEquation} without projection. As a result, the solution is evolving along the trajectories of the grid equation~\eqref{eq:InvariantGridEquation}. Since the spacing between trajectories is not constant, i.e.\ $h^{+}$ and $h^{-}$ are changing in time, we choose to plot the error versus $1/N$. In Fig.~\ref{fig:convergence_no_remapping}, we observe the second order convergence expected.

\subsection{Invariant scheme with non-invariant quadratic interpolation}

\begin{figure}[ht!]
\centering
\includegraphics[scale=0.38]{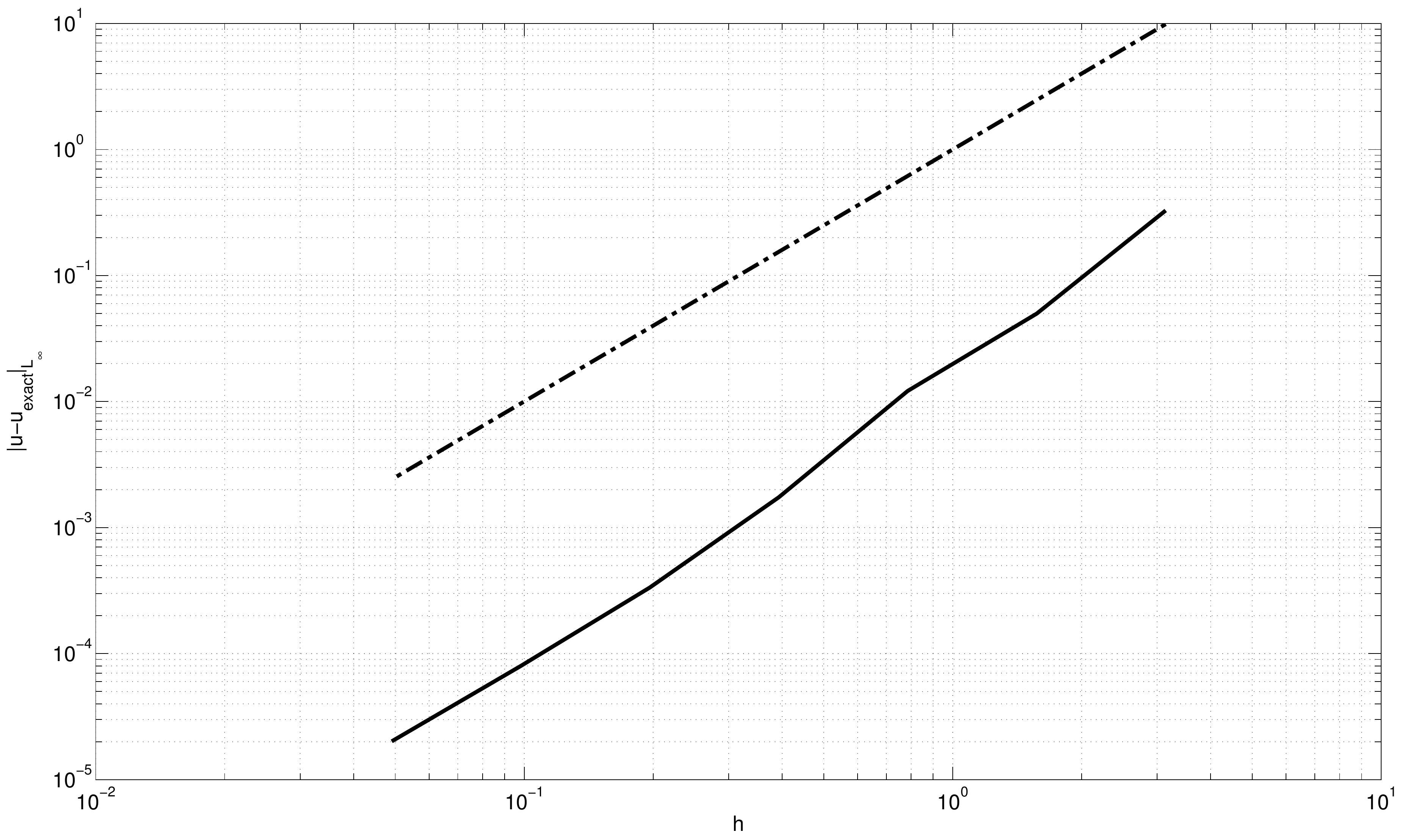}
\caption{Convergence plot for the invariant scheme~\eqref{eq:InvariantSchemeHeatEquation} with invariant grid equation~\eqref{eq:InvariantGridEquation} using non-invariant quadratic interpolation~\eqref{eq:QuadraticInterpolationNonInvariant} as projection.}
\label{fig:convergence_non-invariant_interpolation}
\end{figure}

In this scheme we interpolate the solution of the invariant scheme~\eqref{eq:InvariantSchemeHeatEquation} at every step back onto the regular grid using standard Lagrange quadratic interpolation~\eqref{eq:QuadraticInterpolationNonInvariant}. As a result, we have the solution on a regular grid with step-size $h=1/N$. In Fig.~\ref{fig:convergence_non-invariant_interpolation}, we observe the second order convergence expected.

\subsection{Invariant scheme with invariant quadratic interpolation}

\begin{figure}[ht!]
\centering
\includegraphics[scale=0.38]{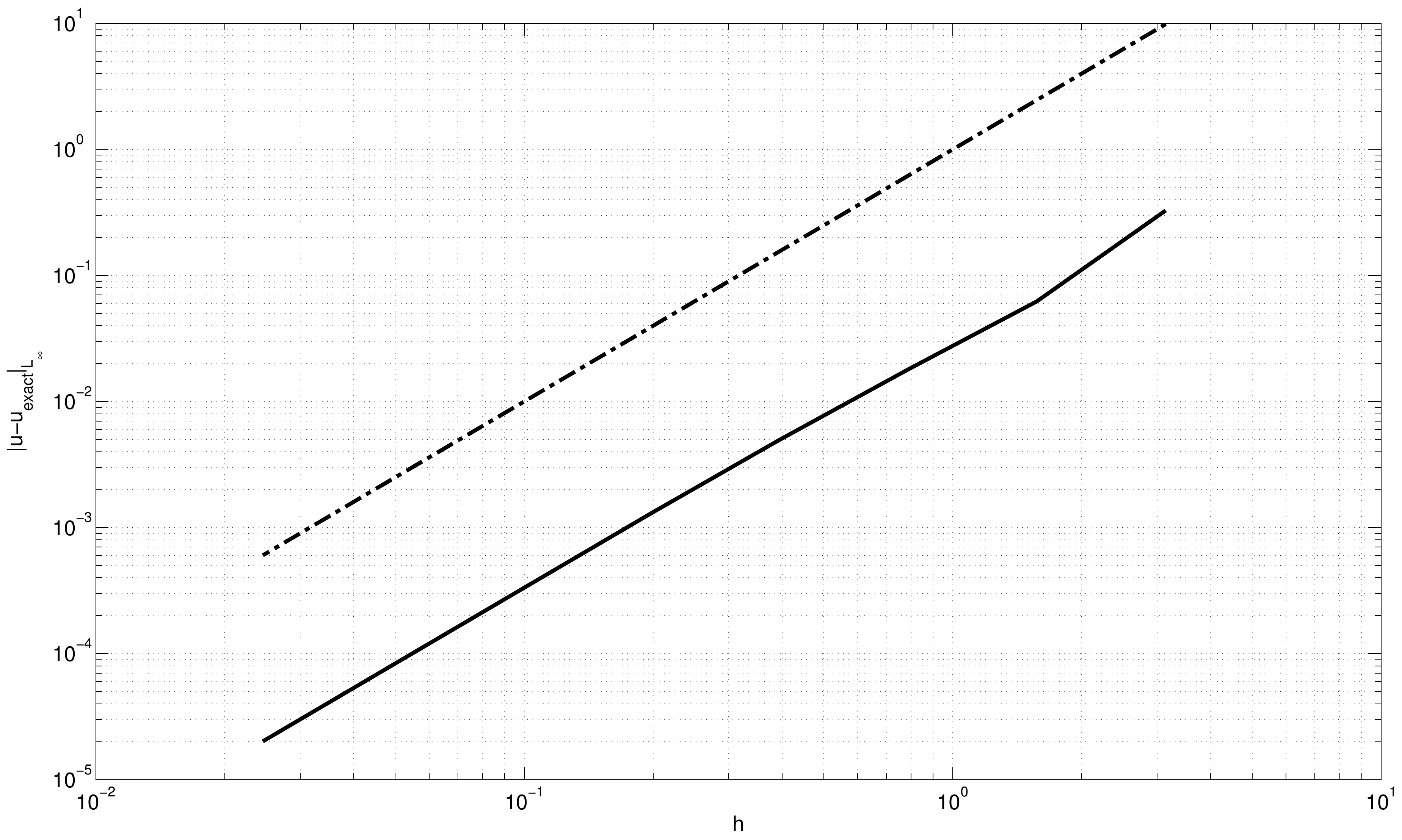}
\caption{Convergence plot for the invariant scheme~\eqref{eq:InvariantSchemeHeatEquation} with invariant grid equation~\eqref{eq:InvariantGridEquation} using invariant quadratic interpolation~\eqref{eq:QuadraticInterpolationInvariant} as projection.}
\label{fig:convergence_invariant_interpolation}
\end{figure}

In this scheme, we interpolate the solution of the invariant scheme~\eqref{eq:InvariantSchemeHeatEquation} at every step back onto the regular grid using the invariant Lagrange quadratic interpolation~\eqref{eq:QuadraticInterpolationInvariant} described in the previous section. As for the case above, at each time step we have the solution on a regular grid with step-size $h=1/N$. In Fig.~\ref{fig:convergence_invariant_interpolation}, we observe the second order convergence expected.

\subsection{Linearity preservation in the invariant numerical scheme}

Linearity is not preserved by construction in the schemes proposed. For this reason it is instructive to check numerically whether or not linearity is preserved in the fully invariant scheme.

Consider the initial value problem,
\begin{gather*}
 u_{t} = u_{xx}\\
 u (x,t=0 )= (\sin(x-1)+2)+(\cos x+2),
\end{gather*}
the solution of which we call $u_{\rm exact}$. We then solve numerically the following two equations
\begin{gather*}
 u_{t}^{a} = u_{xx}^{a}\ \text{with}\  u^{a}(x,0)=\sin(x-1)+2,\\
 u_{t}^{b} = u_{xx}^{b}\ \text{with}\  u^{b}(x,0)=\cos x+2,
\end{gather*}
and def\/ine $u_{s}=u^{a}+u^{b}.$

\begin{figure}[ht!]
\centering
\includegraphics[scale=0.38]{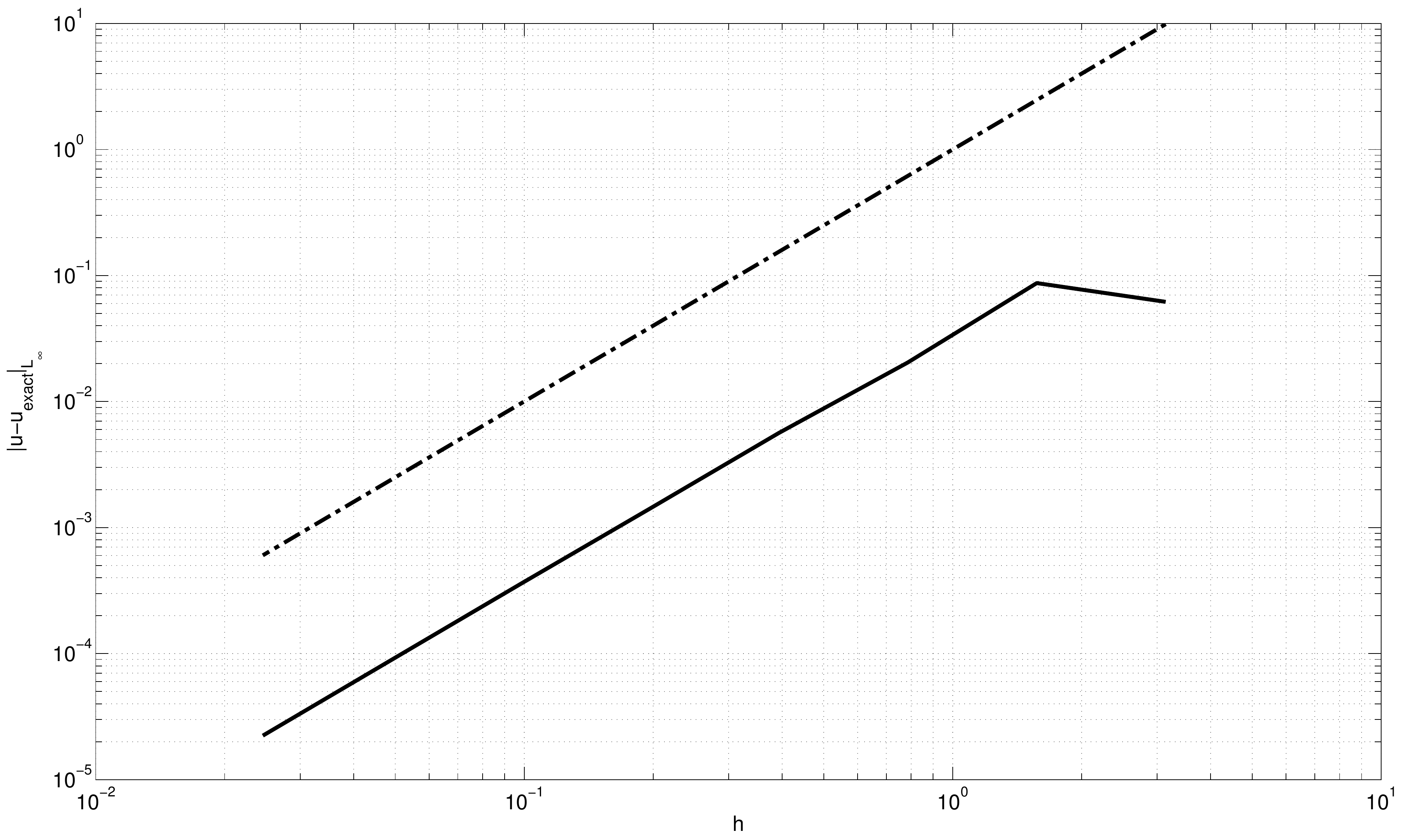}
\caption{Convergence plot for the linearity test using the invariant scheme~\eqref{eq:InvariantSchemeHeatEquation} with invariant grid equation~\eqref{eq:InvariantGridEquation} using invariant quadratic interpolation~\eqref{eq:QuadraticInterpolationInvariant} as projection.}
\label{fig:convergence_linearity}
\end{figure}

Fig.~\ref{fig:convergence_linearity} depicts the $L_\infty$ error between $u_{\rm exact}$ and $u_s$. We observe a convergence rate of second order. In other words, despite the fully invariant numerical scheme does not explicitly preserve the symmetry associated with the linear superposition principle, we observe that the linearity property is preserved approximately to the order of the method.

\section{Conclusions}\label{sec:ConculsionsHeatEquation}

In this paper we construct invariant discretization schemes using the method of invariantization via equivariant moving frames. The advantage of this technique is that it allows one to start with a given non-invariant scheme and convert this initial scheme into a f\/inite dif\/ference approximation of a system of dif\/ferential equations $\mathcal L$ that is invariant under the same maximal Lie invariance group $G$ (or a suitably chosen subgroup of $G$) as admitted by $\mathcal L$.

The possibility of converting non-invariant numerical schemes into invariant discretizations may lead to the overly optimistic speculation that the schemes constructed by invariantization could be easily included in existing numerical models using the original scheme. The hurdle preventing this in practice is that preserving symmetry groups of systems of evolution equations more complicated than scalings or translations requires the use of moving grids. Converting numerical models that use standard discretization schemes based on f\/ixed lattices to (invariant) discretization schemes on moving meshes is not an easy task. At the same time, rewriting numerical models from scratch for the simulation of involved physical processes using symmetry-preserving schemes might be a time-consuming and costly task too and it not certain that this is feasible at all. Moreover, it is as of now unclear whether preserving symmetries in numerical schemes for multi-dimensional systems of partial dif\/ferential equations gives enough added value compared to standard schemes that one might justify such an undertaking in practice.

This is why one relies on f\/inding methods allowing one to ef\/f\/iciently include invariant discretization schemes into existing numerical models without the need to rewrite new models from scratch that incorporate the invariance methodology. The method proposed in this article solves this problem by breaking the integration procedure into two steps, the time-stepping using the invariant numerical scheme with an invariant numerical grid equation and the projection (interpolation) of the results obtained at intermediate grid points to the regular mesh. This interpolation can be done in an invariant way by applying the moving frame map used to invariantize the initial discretization scheme also to a particular interpolation method. An alternative is to assemble the invariant interpolation method using joint invariants. Either way, it is worthwhile pointing out that interpolations requiring only data given at a single time level are already invariant under most symmetry groups as admitted by physical systems of dif\/ferential equations. Thus, invariantization of interpolation formulas will often only lead to minor modif\/ications of the initial interpolation method chosen and the inf\/luence on the numerical solution might be rather small. In the numerical tests carried out above for the heat equation, the dif\/ference in the convergence properties we found when using invariant or non-invariant interpolation methods is indeed small although using the invariant interpolation gave slightly better numerical results. This is encouraging and the reason why we plan to further investigate invariant numerical schemes using the projection procedure.

We illustrate the evolution--projection strategy by integrating the one-dimensional linear heat equation with an invariant numerical scheme. The heat equation has been studied quite extensively in light of its invariance properties and in particular it is a standard model for the construction of invariant numerical schemes~\cite{baki97Ay,doro03Ay,vali05Ay}. At the same time, a comprehensive numerical analysis of such schemes was not given before and thus seems relevant to be reported. This is another aim of the present paper. Again, the analysis of numerical properties of discretization schemes is considerably easier if one can use non-evolving meshes.

Further work we intend to do is to employ the evolution--projection strategy to multi-dimensional systems of dif\/ferential equations using both higher-order discretization and interpolation schemes.

\subsection*{Acknowledgements}

The authors thank Professor Roman Popovych for valuable discussions and careful reading of the manuscript. The valuable remarks of the anonymous referees are much appreciated. This research was supported by the Austrian Science Fund (FWF), project J3182--N13 (AB). JCN wishes to acknowledge partial support from the NSERC Discovery Program, and the National Science Foundation through grant
DMS-0813648.

\pdfbookmark[1]{References}{ref}
\LastPageEnding

\end{document}